\documentclass[12pt,preprint]{aastex}

\newcommand{\msun}{\mbox{$M_{\odot}$}}
\newcommand{\Msun}{\mbox{$M_{\odot}$}}

\newcommand{\vinf}{\mbox{$v_{\infty}$}}

\newcommand{\mdot}{\mbox{$\dot{M}$}}

\newcommand{\msunyr}{\mbox{$M_{\odot} {\rm yr}^{-1}$}}

\newcommand{\factor}{f}

\begin{document}

\title{The transition mass-loss rate: Calibrating the role of line-driven winds in massive star evolution}
\author{Jorick S. Vink$^1$ and G{\"o}tz Gr{\"a}fener$^1$}
\affil{$^!$Armagh Observatory, College Hill, BT61 9DG Armagh, Northern Ireland, UK; jsv@arm.ac.uk}

\begin{abstract}
A debate has arisen regarding the importance of stationary
versus eruptive mass loss for massive star evolution. 
The reason is that stellar winds have been found to be clumped, which results 
in the reduction of unclumped empirical mass-loss rates. 
Most stellar evolution models employ theoretical mass-loss rates which are 
{\it already} reduced by a moderate factor of $\simeq$2-3 compared to non-corrected empirical rates. 
A key question is whether these reduced rates are 
of the correct order of magnitude, or if they should be reduced even further, which
would mean that the alternative of eruptive mass loss becomes necessary.  
Here we introduce the transition mass-loss rate $\dot{M}_{\rm trans}$
between O and Wolf-Rayet (WR) stars.
Its novelty is that it is model independent. All that is required is postulating
the {\it spectroscopic} transition point in a given data-set, and determining the 
stellar luminosity, which is far less model dependent than the 
mass-loss rate. The transition mass-loss rate 
is subsequently used to calibrate stellar wind strength by its application 
to the Of/WNh stars in the Arches cluster. 
Good agreement is found with two alternative modelling/theoretical results, 
suggesting that the rates provided by current theoretical models 
are of the right order of magnitude in the $\sim$50\Msun\ mass range.
Our results do not confirm the specific {\it need} for eruptive mass loss as  
Luminous Blue Variables, and current stellar evolution modelling 
for Galactic massive stars seems sound.
Mass loss through alternative mechanisms might still become necessary 
at lower masses, and/or metallicities, and the {\it quantification} of 
alternative mass loss is desirable.  
\end{abstract}


\section{Introduction}
\label{s_intro}

Mass loss via stellar winds is thought to play a dominant role 
in the evolution of massive O-type stars, because of the loss of {\it mass}, as winds 
``peel off'' the star's outermost layers (Conti 1976), as well as through 
the associated loss of {\it angular momentum} (e.g. Langer 1998, Meynet \& Maeder 2002). However, during
the last decade, large uncertainty has been pointed out regarding our quantitative 
knowledge of the mass-loss rates of massive stars, as 
stellar winds have been revealed to be clumped, resulting in empirical rates that have been 
overestimated. 

Although it had been known for decades that O-type winds are clumped 
(Lupie \& Nordsieck 1987, Eversberg et al. 1998), the severity 
did not appear to be fully recognized until Bouret et
al. (2005) and Fullerton et al. (2006) claimed mass-loss reductions
of factors $\sim$3-7 and $\sim$20-130 respectively in comparison to unclumped
H$\alpha$ and radio mass-loss rates (e.g. Lamers \& Leitherer 1993).
The H$\alpha$ diagnostics depends on the density squared, and are thus
sensitive to clumping, whilst ultraviolet P Cygni lines such as 
P{\sc v} are {\it in}sensitive to clumping as these depend linearly on 
the density. The above-mentioned Bouret et al. and Fullerton et al.  analyses were
based on models where the wind is divided into a portion of the wind
containing all the material with a volume filling factor $f_{\rm V}$
(the reciprocal of the clumping factor), whilst the remainder of the
wind is assumed to be void. This pure {\it micro}-clumping approach
is probably an oversimplification of the real situation, but it
provides interesting insights into the potential mass-loss rate
reductions.

In reality, clumped winds are likely porous, with a range of clump
sizes, masses, and optical depths. {\it Macro}-clumping and porosity
have been investigated with respect to both the spectral analyses
(e.g. Oskinova et al. 2007, Sunqvist et al. 2010, Surlan et al. 2012) 
as well as the
radiative driving (Muijres et al. 2011). The upshot from these studies
is that O star mass-loss rates may only be reduced by a moderate factor
of $\sim$3 (Repolust et al. 2004, 
Puls et al. 2008), which would bring their clumping properties 
in agreement with those of Wolf-Rayet (WR) winds, for which similar 
moderate clumping factors have been derived (Hamann \& Koesterke 1998).
The latter are based on the analysis of emission line wings due to
electron scattering, which have the advantage that they do not
depend on detailed ionization fractions and abundances of trace
elements. These moderate clumping factors would imply that massive star
evolution modelling is not affected, as current state-of-the-art
rotating stellar models (e.g. Georgy et al.  2011; Brott et al. 2011) already employ
moderately reduced rates via the theoretical relations of Vink et al. (2000).

In light of the severe mass-loss reductions claimed e.g. by Fullerton 
et al. (2006), Smith \& Owocki (2006) argued that the
integrated mass loss from stationary stellar winds for very massive
stars (VMS) above $\simeq$50\msun\ may be vastly insufficient to explain
their role as the progenitors of WR stars and stripped-envelope Ibc
supernovae. Instead, Smith \& Owocki argued that the bulk of VMS mass loss 
is likely of an eruptive rather than a stationary nature. In
particular, they highlighted the alternative option of eruptive mass 
loss during the Luminous Blue Variable (LBV) phase. 

In view of the new porosity results, the arguments of Smith \&
Owocki (2006) however seem to have lost weight. 
Furthermore, quantitative estimates on the integrated amount 
of eruptive mass loss are hard to obtain as both the eruption frequency, and 
the amounts of mass lost per eruption span a wide range (of a
factor 100) with LBV nebular mass estimates varying from 
$\sim$0.1\msun\ in P\,Cygni to $\sim$10\msun\ in $\eta$\,Car, as discussed by 
Smith \& Owocki (2006). Moreover, 
the energies required to produce such giant mass
eruptions are very high ($\simeq 10^{50}$\,erg), and their energy
source is unknown. Soker (2004) discussed that the energy
and angular momentum required for $\eta$\,Car great eruption
cannot be explained within a single-star scenario. 

Whilst stationary winds in O and WR stars are ubiquitous, it is 
not at all clear if LBV-type
objects like $\eta$\,Car have encountered a special evolution (such as a
merger) or if all massive stars go through eruptive mass-loss phases. 
On the other hand, for the most massive main-sequence WNh stars
(Crowther et al. 2010, Bestenlehner et al. 2011) there is both theoretical and 
empirical evidence for strong Eddington parameter $\Gamma$-dependent (see definition 
Eq.\,\ref{eq_gamma}) mass loss (Gr\"afener et al.  2011). 
For VMS the role of stationary mass loss has thus increased rather
than decreased in recent years.

In summary, the relevant roles of eruptive versus stationary mass loss 
seem rather uncertain and unsettled at the current time. There are ongoing debates
as to whether wind clumping reduces the mass-loss rates by moderate factors of $\simeq$2-3, such 
that stellar evolution would not be affected, or by more severe factors of order $\sim$10. 
In the latter case, line-driving 
would become negligible and alternatives such as 
eruptive mass loss would need to be considered. 

In order to address the relative role of wind versus eruptive mass loss,
it would be beneficial to be able to calibrate either one of them.
At the moment both stellar wind and eruptive mass loss 
could be inaccurate by factors of 10, and possibly even more. 
In this Letter, we attempt to alleviate this problem by presenting a 
methodology that involves a model-independent mass-loss indicator, the transition mass-loss 
rate $\dot{M}_{\rm trans}$ -- located 
right at the transition from optically thin to optically thick stellar winds. 
Martins et al. (2008) found two mass-loss relations for VMS Arches cluster stars, one 
for the Of stars and one for the late-type 
WNh stars respectively.
The fact that Wolf-Rayet stars with WNh spectral classification
have optical depth larger than one has already been discussed in  
literature (e.g. Gr\"afener \& Hamann 2008), and one 
might thus expect to witness a transition from optically thin 
O-type winds to optically thick Wolf-Rayet winds.

Vink et al. (2011) discovered a sudden change in the slope of the
  mass-loss versus $\Gamma$ relation at the transition from O-type
  (optically thin) to WR-type (optically thick) winds. 
Interestingly, this transition
  was found to occur for a wind efficiency parameter $\eta$ $=$
  $\mdot$ $\vinf$/($L/c$) of order unity. This key result 
from Monte Carlo modelling that the transition from O to WR-type mass loss
  coincides with $\eta \simeq 1$, can also be found analytically
  (Sect.\,\ref{sec_anal}). 
And the result can be utilized to
``calibrate'' wind mass loss in an almost model independent manner
(Sect.\,\ref{sec_arches}).

\section{The transition mass-loss rate}
\label{sec_anal}

Netzer \& Elitzur (1993) 
and Lamers \& Cassinelli (1999; hereafter LC99) give 
general momentum considerations for dust-driven winds 
(see LC99 pages 152-153) that can also be applied to line-driven winds. 
The integral form of the momentum equation contains four terms (Eq. 7.5 of LC99). 
Because hydrostatic equilibrium is a good approximation 
for the subsonic part of the wind, and the gas pressure 
gradient is small beyond the sonic point, LC99 
argue that the second and third terms are negligible 
compared to the first and fourth, resulting in:

\begin{equation}
\int_{R_{\star}}^{\infty} 4 \pi r^2 \rho v \frac{dv}{dr} dr + \int_{r_{\rm s}}^{\infty} \frac{G M}{r^2} (1 - \Gamma)  \rho 4 \pi r^2 dr = 0.
\label{eq_LC99}
\end{equation}

Employing the mass-continuity equation $\dot{M} = 4 \pi r^2 \rho v$, one obtains

\begin{equation}
\int_{R_{\star}}^{\infty} \dot{M} \frac{dv}{dr} dr = \dot{M} v_{\infty} = 4 \pi G M \int_{r_{\rm s}}^{\infty} (\Gamma(r) - 1)\rho dr 
\end{equation}

where $r_s$ denotes the sonic radius and $\Gamma(r)$ the Eddington
factor with respect to the total flux-mean opacity $\kappa_{\rm F}$:

\begin{equation}
\label{eq_gamma}
  \Gamma(r) = \frac{\kappa_{\rm F} L}{4\pi c GM}.
\end{equation}

Using the wind optical depth 
$ \tau = \int_{r_s}^\infty
\kappa_{\rm F}\rho\, {\rm d}r$, one obtains

\begin{equation}
\dot{M} v_{\infty} \simeq \frac{4 \pi G M}{\kappa} (\Gamma - 1) \tau = \frac{L}{c} \frac{\Gamma-1}{\Gamma} \tau.
\label{eq_gaga}
\end{equation}

Where it is assumed that $\Gamma$ is
significantly larger than one, and the factor
$\frac{\Gamma-1}{\Gamma}$ is thus close to unity
(LC99's second assumption), resulting in 

\begin{equation}
\dot{M} v_{\infty} = \frac{L}{c} \tau .
\end{equation}

One can now derive a key condition for the wind efficiency number $\eta$

\begin{equation}
\eta = \frac{\dot{M} v_{\infty}}{L/c} = \tau = 1 .
\label{eq_eta}
\end{equation}

The key point of our Letter is that we can employ the unique condition $\eta = \tau = 1$ 
right at the transition
from optically thin O-star winds to optically-thick WR winds, and obtain a {\it model-independent} $\dot{M}$. 
In other words, if we were to have an empirical data-set available that contains luminosity determinations for 
O and WR stars, we can obtain the transition mass-loss rate $\dot{M}_{\rm trans}$ simply by considering
the transition luminosity $L_{\rm trans}$ and the terminal velocity $v_{\infty}$
representing the transition point from O to WR stars:

\begin{equation}
\dot{M}_{\rm trans} = \frac{L_{\rm trans}}{v_{\infty} c}
\label{eq_transm}
\end{equation}

We note that this transition point can be obtained by purely spectroscopic means, 
{\em independent} of any assumptions regarding wind clumping.

\subsection{Testing the assumptions}
\label{sec_mod}

\begin{table}
\label{tab_models}
\centering
\caption{Wind models. If not stated otherwise, the adopted stellar parameters
     are $T_\star=35$\,kK and $\log(L/L_\odot)=6.0$ (with $M=60\,M_\odot$).}
\small{\begin{tabular}{lllllll}  \hline 
\hline 
    \rule{0cm}{2.2ex}$\beta$ & $\log(\dot{M})$ & $v_\infty$ & $\eta$ & $\eta'$ & $\tau$ & $\factor$ \\
    &  [$\frac{M_\odot}{\rm yr}$]  &  [$\frac{\rm km}{\rm s}$]  &  &  & \\
    \hline
    \multicolumn{7}{l}{\rule{0cm}{2.2ex}WR\,22, $M=78.1\,M_\odot$, $\log(L/L_\odot)=6.3$} \\                                     
    \hline \rule{0cm}{2.2ex}%
    HYD  & -4.868 &  979.5 &  0.326 &  0.315 &  1.198 &  0.272 \\
    1.0  & -4.868 &  979.5 &  0.326 &  0.310 &  1.160 &  0.281 \\
    1.0  & -4.607 & 1785.0 &  1.085 &  1.062 &  1.968 &  0.551 \\
    \hline
    \hline
    \multicolumn{7}{l}{\rule{0cm}{2.2ex}
      $v_\infty/v_{\rm esc}=2.5 \rightarrow (\Gamma-1)/\Gamma \sim 2.5/3.5 = 0.714$}\\
    \hline \rule{0cm}{2.2ex}%
    0.5  & -4.905 & 2289.6 &  1.4 &  1.385 &  1.681 &  0.833 \\
    0.5  & -5.148 & 2289.6 &  0.8 &  0.793 &  0.956 &  0.837 \\
    0.5  & -5.750 & 2289.6 &  0.2 &  0.200 &  0.238 &  0.840 \\
    \hline \rule{0cm}{2.2ex}%
    1.0  & -4.905 & 2289.6 &  1.4 &  1.383 &  1.938 &  0.722 \\
    1.0  & -5.148 & 2289.6 &  0.8 &  0.793 &  1.091 &  0.733 \\
    1.0  & -5.750 & 2289.6 &  0.2 &  0.198 &  0.280 &  0.713 \\
    \hline \rule{0cm}{2.2ex}%
    1.5  & -4.905 & 2289.6 &  1.4 &  1.386 &  2.293 &  0.610 \\
    1.5  & -5.148 & 2289.6 &  0.8 &  0.794 &  1.299 &  0.616 \\
    1.5  & -5.750 & 2289.6 &  0.2 &  0.199 &  0.330 &  0.605 \\
    \hline
    \multicolumn{7}{l}{\rule{0cm}{2.2ex}
    $v_\infty/v_{\rm esc}=1.5 \rightarrow (\Gamma-1)/\Gamma \sim 1.5/2.5 = 0.600$}\\
    \hline \rule{0cm}{2.2ex}%
    0.5  & -4.683 & 1373.8 &  1.4 &  1.365 &  2.046 &  0.684 \\
    0.5  & -4.926 & 1373.8 &  0.8 &  0.780 &  1.179 &  0.678 \\
    0.5  & -5.528 & 1373.8 &  0.2 &  0.195 &  0.301 &  0.664 \\
    \hline \rule{0cm}{2.2ex}%
    1.0  & -4.683 & 1373.8 &  1.4 &  1.362 &  2.644 &  0.529 \\
    1.0  & -4.926 & 1373.8 &  0.8 &  0.780 &  1.531 &  0.522 \\
    1.0  & -5.528 & 1373.8 &  0.2 &  0.196 &  0.388 &  0.516 \\
    \hline \rule{0cm}{2.2ex}%
    1.5  & -4.683 & 1373.8 &  1.4 &  1.364 &  3.231 &  0.433 \\
    1.5  & -4.926 & 1373.8 &  0.8 &  0.779 &  1.971 &  0.406 \\
    1.5  & -5.528 & 1373.8 &  0.2 &  0.196 &  0.490 &  0.408 \\
    \hline
    \hline
    \multicolumn{7}{l}{\rule{0cm}{2.2ex}
      $v_\infty/v_{\rm esc}=2.5$, $T_\star=70$\,kK}  \\
    \hline \rule{0cm}{2.2ex}%
    0.5  & -5.449 & 4579.2 &  0.8 &  0.800 &  0.955 &  0.837 \\
    1.0  & -5.449 & 4579.2 &  0.8 &  0.800 &  1.140 &  0.701 \\
    1.5  & -5.449 & 4579.2 &  0.8 &  0.800 &  1.442 &  0.555 \\
    \hline
    \multicolumn{7}{l}{\rule{0cm}{2.2ex}
      $v_\infty/v_{\rm esc}=2.5$, $T_\star=17.5$\,kK}  \\
    \hline \rule{0cm}{2.2ex}%
    0.5  & -4.847 & 1144.8 &  0.8 &  0.784 &  0.946 &  0.846 \\
    1.0  & -4.847 & 1144.8 &  0.8 &  0.784 &  1.033 &  0.774 \\
    1.5  & -4.847 & 1144.8 &  0.8 &  0.783 &  1.203 &  0.665 \\
    \hline
    \multicolumn{7}{l}{\rule{0cm}{2.2ex}
      $v_\infty/v_{\rm esc}=2.5$, $\log(L/L_\odot)=6.3$ (with $M=120\,M_\odot$)}  \\
    \hline \rule{0cm}{2.2ex}%
    0.5  & -4.923 & 2724.4 &  0.8 &  0.794 &  0.959 &  0.833 \\
    1.0  & -4.923 & 2724.4 &  0.8 &  0.794 &  1.121 &  0.714 \\
    1.5  & -4.923 & 2724.4 &  0.8 &  0.795 &  1.366 &  0.585 \\
    \hline
    \multicolumn{7}{l}{\rule{0cm}{2.2ex}
      $v_\infty/v_{\rm esc}=2.5$, $\log(L/L_\odot)=5.7$ (with $M=40\,M_\odot$)}  \\
    \hline \rule{0cm}{2.2ex}%
    0.5  & -5.438 & 2237.6 &  0.8 &  0.794 &  0.953 &  0.839 \\
    1.0  & -5.438 & 2237.6 &  0.8 &  0.794 &  1.085 &  0.737 \\
    1.5  & -5.438 & 2237.6 &  0.8 &  0.794 &  1.276 &  0.627 \\
    \hline \end{tabular}
} 
\end{table}

In the above analysis we made two assumptions that we wish to 
check with numerical tests involving sophisticated 
hydrodynamic wind models (from Gr\"afener \& Hamann 2008) 
and simpler $\beta$-type velocity laws, commonly used in O/WR wind modelling. 
The results are compiled in Tab.\,1.
We first confirm LC99's first assumption through the 
comparison of $\eta$, determined from $\dot{M}$ and $v_\infty$, to 
the approximate $\eta'$ values as computed from the right-hand-side of integral 
in Eq.\,\ref{eq_LC99}, where $\eta' L/c = 4 \pi G M \int_{r_{\rm s}}^{\infty} (\Gamma(r) - 1) \rho dr$. 
Evidently, the values of $\eta$ and $\eta'$
agree at the few percent level, and the first LC99 
approximation is verified.

Second, we investigate the assumption that the term
$\frac{\Gamma - 1}{\Gamma}$ in Eq.\,\ref{eq_gaga} is close to
unity by numerical integration of $\tau =
\int_{r_s}^\infty \kappa\rho\, {\rm d}r$.  We obtain a
correction factor $\factor$, which we define by
\begin{equation}
  \dot M v_\infty
  \equiv  \factor \frac{L}{c} \tau.
\end{equation}
With this definition Eq.\,\ref{eq_eta} becomes
\begin{equation}
  \eta = \frac{\dot M v_\infty}{L/c} = \factor \tau.
\end{equation}

To compute the integral numerically, we need to obtain the 
density $\rho(r)$, and the flux-mean opacity $\kappa_{\rm F}(r)$
in the stellar wind ($\Gamma(r)$ follows from
Eq.\,(\ref{eq_gamma}). The hydrodynamic wind models of
Gr\"afener \& Hamann (2008) have these quantities directly available. 
We have performed a direct computation for the first model (HYD) 
in Tab.\,1 for the Galactic WNh star WR\,22
(Gr\"afener \& Hamann 2008). 
For this model we obtain $\factor=0.272$.
This value is lower than, but of the order of, unity.  
The terminal wind speed in this model is significantly 
lower than the observed value for WR\,22 (980\,km/s vs. 1785\,km/s). 
Consequently, our derived $\factor$ is likely on the low side. 
We expect $\Gamma = g_{\rm rad}/g$ to be connected to the ratio
$(v_\infty+v_{\rm esc})/v_{\rm
  esc}=v_\infty/v_{\rm esc}+1$, and $\factor \simeq
\frac{\Gamma-1}{\Gamma}$.

Here we 
follow a model-independent approach, adopting $\beta$-type velocity laws.
The mean opacity $\kappa_F$ then follows from the resulting
radiative acceleration $g_{\rm rad}$ 
\begin{equation}
g_{\rm rad}(r) = \kappa_F(r)\frac{L}{4\pi c r^2}.
\end{equation}
$g_{\rm rad}$ follows from the prescribed density $\rho(r)$ and velocity
structures $v(r)$ via the equation of motion
\begin{equation}
  v \frac{{\rm d}v}{{\rm d}r} = g_{\rm rad} - \frac{1}{\rho}\frac{{\rm d}p}{{\rm d}r}
-\frac{GM}{r^2},
\end{equation}
where we assume a grey temperature structure to compute the gas
pressure $p$. We note that these results are completely independent 
of any assumptions regarding wind porosity, or the chemical composition 
of the wind material. The {\em only}
  assumption that goes into these considerations is that the winds
  are radiatively driven. The resulting mean opacity $\kappa_F$
  consequently captures all physical effects that could potentially affect
  the radiative driving.

The obtained values for the correction factor $\factor$ are summarized
in Tab.\,1. The first three models in
Tab.\,1 represent a consistency test with the
hydrodynamic model for WR\,22. Using a beta law with $\beta=1$, 
and the same $v_\infty$, we obtain almost
exactly the same $\factor$ as for the hydrodynamic model, justifying 
our $\beta$-law approach, which we employ in the following.
Now employing the {\em observed} -- and therefore likely close to correct -- 
value of $v_\infty$ (and a correspondingly increased $\dot{M}$, we obtain 
$\factor = 0.55$. 

To get a handle on the overall behaviour of this factor $\factor$, 
we computed a series of wind models for a range of stellar parameters 
$17.5<T_\star/{\rm kK}<70\,$ and $5.7<\log(L/L_\odot)<6.3$, 
with wind efficiencies around the transition region ($0.2<\eta<1.4$). 
Remarkably, the resulting values of $\factor$ 
depend {\em only} on the adopted values of $v_\infty/v_{\rm esc}$ 
and $\beta$. For $v_\infty/v_{\rm esc}=2.5$, we obtain $\factor\sim0.8,$ 0.7,
0.6 respectively for $\beta=0.5$, 1.0, 1.5, where the last value
is probably most appropriate (Vink et al. 2011).   
Overall, we derive values of $f$ in the range 0.4-0.8, 
with a mean value of 0.6. We note that the error on this number $f$ 
is within the uncertainty of the luminosity determinations described in the next 
section.

For transition objects with $\tau\simeq 1$ we thus expect
that $\factor \simeq0.6$, i.e.\ the transition between O and WR
spectral types should occur at mass-loss rates of
\begin{equation}
  \dot{M} = \factor \frac{L_{\rm trans}}{v_{\infty} c} \simeq 0.6 \dot{M}_{\rm trans}.
\end{equation}
The fact that the correction factor is within a factor of two 
of our idealized approach (Eq.~\ref{eq_transm}) is highly encouraging. 
We stress that this number is independent of any potential model deficiencies, 
as we have used the observed values of $v_\infty$ in this analysis.

\section{The transition mass loss rate in the Arches cluster}
\label{sec_arches}

\begin{table} 
\label{tab_arches}
  \centering
\caption{The transition from O to WR stars for the most
    massive stars in the Arches Cluster.}
\begin{tabular}{llllllc} \hline \hline
      \rule{0cm}{2.2ex}Star & subtype & $\log(L)$   & $\log(\dot{M})$  & $v_\infty$ & $\log(\dot{M}_{\eta=1})$ \\
      &       & [$L_\odot$] & [$\frac{M_\odot}{\rm yr}$] & [$\frac{\rm km}{\rm s}$] & [$\frac{M_\odot}{\rm yr}$] \\
      \hline \rule{0cm}{2.2ex}%
      F9   &WN8-9   &6.35 &$-$4.78 &1800 &$-$4.60   \\
      F1   &WN8-9   &6.30 &$-$4.70 &1400 &$-$4.54   \\
      F14  &WN8-9   &6.00 &$-$5.00 &1400 &$-$4.84   \\
      B1   &WN8-9   &5.95 &$-$5.00 &1600 &$-$4.95   \\
      F16  &WN8-9   &5.90 &$-$5.11 &1400 &$-$4.94   \\
      \hline \rule{0cm}{2.2ex}%
      F15  &O4-6If+ &6.15 &$-$5.10 &2400 &$-$4.92 & \\
      F10  &O4-6If+ &5.95 &$-$5.30 &1600 &$-$4.95 & $\dot{M}_{\rm trans}$\\
      \hline \rule{0cm}{2.2ex}%
      F18  &O4-6I   &6.05 &$-$5.35 &2150 &$-$4.98   \\
      F21  &O4-6I   &5.95 &$-$5.49 &2200 &$-$5.09   \\
      F28  &O4-6I   &5.95 &$-$5.70 &2750 &$-$5.18   \\
      F20  &O4-6I   &5.90 &$-$5.42 &2850 &$-$5.25   \\
      F26  &O4-6I   &5.85 &$-$5.73 &2600 &$-$5.26   \\
      F32  &O4-6I   &5.85 &$-$5.90 &2400 &$-$5.22   \\
      F33  &O4-6I   &5.85 &$-$5.73 &2600 &$-$5.26   \\
      F22  &O4-6I   &5.80 &$-$5.70 &1900 &$-$5.17   \\
      F23  &O4-6I   &5.80 &$-$5.65 &1900 &$-$5.17   \\
      F29  &O4-6I   &5.75 &$-$5.60 &2900 &$-$5.41   \\
      F34  &O4-6I   &5.75 &$-$5.77 &1750 &$-$5.19   \\
      F40  &O4-6I   &5.75 &$-$5.75 &2450 &$-$5.33   \\
      F35  &O4-6I   &5.70 &$-$5.76 &2150 &$-$5.33   \\
      \hline \end{tabular}   
 \begin{list}{}{}
 \item[Notes --]Designations,
      subtypes, luminosities ($L$),
      mass-loss rates ($\dot{M}$), and terminal wind velocities
      ($v_\infty$) according to Martins et al. (2008). 
      The 6th column indicates the mass-loss rate where $\eta$=1. 
      For the Arches cluster, we obtain $\log(\dot{M}_{\rm trans})\simeq-4.95$.
\end{list} 
\end{table} 

Martins et al. (2008) analyzed 28 VMS in the Arches cluster, 
with equal numbers of O-type supergiants and nitrogen-rich Wolf-Rayet (WNh) stars 
(sometimes called ``O stars on steroids''). For the O-type supergiants, we 
expect the winds to be optically thin, whilst the WNh stars should have optically thick winds. 
Here we postulate that the O4-6If$+$ represent the transition point where the optical
depth crosses unity. 

In Tab.\,2, we compiled a subset of 20 stars, skipping 
those objects with a He-enriched surface composition. 
The objects are sorted with respect to their spectral subtypes, and within each subtype 
bin with respect to their luminosity. 
Together with the basic stellar and wind parameters
derived by Martins et al. (2008), we list 
the mass-loss rate for which 
$\dot{M}_{\eta = 1}$ if the stars would have a wind
efficiency of exactly 1.
The values listed in Tab.\,2 show that there is 
a transition between O and WR spectral types. 
The {\em spectroscopic} transition for spectral
  subtypes O4-6If+ occurs at  $\log(L)=6.05$ and
  $\log(\dot{M}_{\eta=1}/\msunyr)=-4.95$.

This is the resulting transition mass-loss rate for the Arches cluster stars. 
Its determined value does not depend 
on model uncertainties involving issues such as wind clumping. 
The only remaining uncertainties are due to 
uncertainties in the terminal velocity and the stellar luminosity $L$.  
The latter results from errors in the
distances and reddening parameters, as well as the determination of
effective temperatures from non-LTE model atmospheres. If the
  derived value for $\dot{M}_{\rm trans}$ is compared with empirically
  determined mass-loss rates, the uncertainties in distances and
  reddening parameters nearly cancel, as for empirical mass-loss
  rates based on recombination line analyses $\dot{M}\propto L^{3/4}$,
  while $\dot{M}_{\rm trans}\propto L$. 

To estimate uncertainties in
  the effective temperature scale for O stars we can use historical
  values from the last four decades (e.g. Panagia et al. 1973, Martins
  et al.  2005) as an indicator for potential systematic errors in
the inclusion/neglect of certain micro-physics (line blanketing, wind
effects, etc.), the best error estimate is $\sim$10\% in effective
temperature, leading to potential errors in the luminosity of at most
$\sim$40\%.  This is several factors smaller than the order-of-magnitude
uncertainties in mass-loss rates due to clumping and porosity.  In
other words our simple equation (Eq.\ref{eq_transm}) is of tremendous
value in calibrating stellar wind mass loss, and assessing its role in
the mass loss during the evolution of massive stars
(Sect.\,\ref{sec_disc}).

How does our determination of the transition mass-loss rate compare to other  
methods? Let us first compare our transition mass-loss value to the mass-loss rates of Martins et al. 
Martins et al. use the non-LTE {\sc cmfgen} code by Hillier \& Miller (1998), employing 
a micro-clumping approach with a volume filling factor $f = 0.1$ for their K-band 
analysis.  Their values are in good agreement 
with our transition mass-loss rates for the objects 
at the boundary between O and WR (see their Table 2). 
This is unlikely to be a coincidence. 
According to our findings in Sect.\,\ref{sec_mod}, we expect
  mass-loss rates of the order of $\dot{M} \simeq
  0.6\times\dot{M}_{\rm trans}$ for the transition objects,
  i.e.\ $\log(\dot{M}/\msunyr)=-5.2$.

We also compare the transition mass-loss rate to the 
oft-used theoretical mass-loss relation of Vink et al. (2000), for which we find 
$\log \dot{M}_{\rm Vink} = -5.14\,\msunyr$ for an assumed stellar mass $M = 60\,\msun$.
This number is within 0.2 dex from the transition mass-loss rate $\log \dot{M}_{\rm trans} = -4.95\,\msunyr$. 
The comparison is hardly compromised as a result of the Vink et al. dependence on 
stellar mass, as for masses in the range 40-80\,$\msun$, the Vink et al. 
mass-loss rate varies by at most 0.04 dex. 

In summary, we have three independent mass-loss rate determinations that agree within
a factor of two. This means that our concept of the transition
mass-loss rate has indeed been able to test the accuracy of current mass-loss 
estimates by stellar winds. 

\section{Discussion}
\label{sec_disc}

Now that we have calibrated stellar wind mass loss in the high mass and luminosity regime, 
we assess the role of stellar wind mass loss for massive star evolution. 
For a 60\,$\msun$ star, the main sequence lifetime is $\simeq$3 Myrs (e.g. 
Weidner \& Vink 2010 and references therein). With a stationary mass-loss rate of $10^{-5}$
$\msunyr$ as derived for the transition mass-loss rate in Sect.\,\ref{sec_arches}, this 
means such an object will lose $\simeq$30\,\msun, i.e.
half its initial mass, already on the main sequence during core hydrogen burning. 
We have not yet addressed the new concept of $\Gamma$ dependent mass loss, nor 
any additional stellar wind mass loss during the subsequent core-helium burning 
WR phase. 
In other words, solar-metallicity 60\,$\msun$ stars are expected to lose the bulk  
of their initial masses through stellar winds, leaving very little (if any) space 
for additional (e.g. eruptive) mass loss. 

It is plausible that the strong-winded VMS remain on the 
blue side of the HR diagram, without ever entering an eruptive Luminous 
Blue Variable (LBV) or Red supergiant (RSG) phase. 
Current wisdom thus suggests that solar metallicity VMS likely  
``evaporate'' primarily as the result of stationary wind mass loss, without 
the {\it necessity} of additional eruptive mass loss. However, finding out if 
eruptive mass loss might play an additional role remains an interesting 
exercise, especially for the lower initial mass and sub-solar metallicity ranges, as 
their story might be expected to be different. Moreover, we know 
$\eta$\,Car analogs and supernova impostors 
exist in external galaxies (e.g. Van Dyck et al. 2005, Pastorello et al. 2010, Kochanek et al. 2012).

Contrary to the most massive stars, stars below $\sim$40\,\msun\ 
likely evolve into the RSG, yellow super/hypergaint, and/or LBV regimes 
of the stellar HR diagram (see e.g. Vink 2009).  
We note that the Vink et al. (2000) 
main-sequence mass-loss rates currently in use in stellar models (e.g. 
Brott et al. 2011) for lower mass ``normal'' 20-60\,$\msun$ O stars are 
also already reduced by a factor 2-3 in comparison to previous unclumped 
empirical rates. There is currently no particular 
reason to assume they are still overestimated\footnote{As recently claimed 
in the context of core-collapse supernovae by Smith et al. (2011). 
However, the references quoted in that paper only involve micro-clumping 
studies, and do not consider subsequent work on macro-clumping, as well as 
several alternative studies that aim to calibrate stellar wind mass loss 
(e.g. Voss et al. 2010).}. The results presented here certainly boost 
confidence in the mass-loss rates currently in use, 
although they remain uncalibrated for the lower mass regime. One should also
realize when working down the mass range, starting from our 60$\msun$ calibrator 
star, that the mass-loss rates drop significantly below $10^{-5}$ $\msunyr$. 
Its effects on stellar evolution remain significant due to
the longer evolutionary timescales for lower mass objects and the fact that
it is the multiplication of the mass-loss rate times the duration that is relevant. 
This is especially relevant for angular momentum evolution, possibly down to stellar 
masses as low as 10-15\,\msun\ (Vink et al. 2010).

\end{document}